\documentclass[12pt]{iopart}

\usepackage{iopams}
\usepackage{cite}
\expandafter\let\csname equation*\endcsname\relax
\expandafter\let\csname endequation*\endcsname\relax
\usepackage[T1]{fontenc}

\usepackage{amsmath}
\usepackage{iopams}
\usepackage{graphicx}
\usepackage[breaklinks=true,colorlinks=true,linkcolor=blue,urlcolor=blue,citecolor=blue]{hyperref}
\usepackage{color}
\usepackage{mathrsfs}

\def\t{\tau}
\def\T{\mathcal{T}}
\def\L{\mathscr{L}}

\def\P{\mathbb{P}}
\def\Rho{\mathcal{R}}

\def\erfc{\mathrm{erfc}}
\def\erfcx{\mathrm{erfcx}}

\def\Im{\mathrm{Im}}

\begin{document}

\title[A molecular relay race]{A molecular relay race: sequential
first-passage events to the terminal reaction centre in a cascade of diffusion
controlled processes}

\author{Denis S Grebenkov$^\dagger$, Ralf Metzler$^\sharp$ \& Gleb Oshanin$^\ddagger$}
\address{$\dagger$ Laboratoire de Physique de la Mati\`{e}re Condens\'ee (PMC),
CNRS, Ecole Polytechnique, Institut Polytechnique de Paris, 91120 Palaiseau, France}
\address{$\sharp$ Institute of Physics and Astronomy, University of Potsdam, 14476
Potsdam-Golm, Germany}
\address{$\ddagger$ Sorbonne Universit\'e, CNRS, Laboratoire de Physique Th\'eorique
de la Mati\`ere Condens\'ee (UMR CNRS 7600), 4 Place Jussieu, 75252 Paris Cedex 05,
France}

\begin{abstract}
We consider a sequential cascade of molecular first-reaction events towards a terminal 
reaction centre in which each reaction step is controlled by diffusive motion of
the particles. The model studied here represents a typical reaction setting encountered in diverse 
molecular biology systems, 
in which, e.g., a signal transduction proceeds via a series of consecutive "messengers":  the first messenger has to find its respective immobile target site triggering a launch of the second messenger, the second messenger seeks its own target site and provokes
a launch of the third messenger and so on, resembling a relay race in human competitions. 
For such a molecular relay race taking place in infinite one-, two- and three-dimensional systems, 
 we find exact expressions for the probability density function of the time instant of the terminal reaction event, conditioned 
 on preceding 
 successful reaction events on an ordered array of target sites.
 The obtained expressions pertain to the most general conditions:  
 number of intermediate stages and the corresponding diffusion coefficients, the sizes of the target sites, the distances between them, as well as their reactivities are arbitrary. 

\end{abstract}


\section{Introduction}

In 1916 Smoluchowski calculated the typical time it takes a diffusing molecule
to locate a reaction centre in a three-dimensional setting \cite{smoluchowski}. 
His rate picture of the diffusion limit to a molecular reaction still remains
a cornerstone of physical chemistry \cite{bergvh,calef,atkins,katja}.
Adam and Delbr{\"u}ck \cite{adam} as well as Eigen \cite{eigen} were the first
to realise that the diffusive search of a transcription factor protein for its
specific binding site on the DNA molecule is more complex than the Smoluchowski
picture, involving intermittent three-dimensional excursions in the bulk volume and
one-dimensional motion along the DNA, thus considerably optimising the search
time. This facilitated diffusion picture was championed by Berg, Winter, and
von Hippel \cite{berg,bvh}. Today significant additional detail is known for
this molecular search process \cite{mirny,olivier,gijs,otto}.

Generally, however, reaction-diffusion processes encountered in
molecular biology follow a far more complicated reaction scenario
(see, e.g., Refs. \cite{alberts, bradshow,snustad}) than in the
Smoluchowski or facilitated diffusion pictures. Instead, in a
well-ordered sequence of reaction pathways resembling a
``relay race'' in human competitions, an initial signal propagates along successive reaction
steps, in which the signal often crosses boundaries impermeable to a
single molecular species. 
The ultimately desired cellular response
thus only occurs when all sequential steps are completed in due
order. In a typical setting, the ``first messenger'', e.g., a hormone
or a neurotransmitter, diffuses in an extracellular medium, searching
for an immobile membrane-spanning receptor on the cellular
membrane. Upon binding to this intermediary target, the first
messenger causes a conformational change of the receptor, affecting
its activity and producing the ``primary effector'', which in turn
stimulates a synthesis of the second messenger. The latter propagates
within the cell itself and activates an intracellular process. In
relaying the signal often additional boundaries have to be crossed,
e.g., the membrane of the endoplasmatic reticulum or the nuclear
membrane.

An even more complicated scenario occurs in the ${\rm IP_3/DAG}$
pathway,\footnote{Here, ${\rm IP_3}$ is the  inositol $1,4,5$-triphosphate,
${\rm PIP2}$ is the phospholipid phosphatidylinositol $4,5$-bisphosphate
while ${\rm DAG}$ is the diacyl glycerol \cite{alberts}.} in which membrane
lipids are converted into intracellular second messengers \cite{REF}. Here, two most
important messengers of this type are produced from ${\rm PIP2}$. This lipid
component is cleaved by phospholipase C, an enzyme activated by certain
G-proteins and by calcium ions, that splits the ${\rm PIP2}$ into two smaller
molecules each acting as second messengers. One of them is ${\rm DAG}$, a
molecule that remains within the membrane and activates protein kinase C, which
phosphorylates substrate proteins in both the plasma membrane and elsewhere. The
other messenger is ${\rm IP_3}$, a molecule that leaves the cell membrane and
diffuses within the cytosol. ${\rm IP_3}$ binds to ${\rm IP_3}$-receptors, channels
that release calcium from the endoplasmic reticulum. Thus, the action of ${\rm
IP_3}$ is to produce a third messenger that triggers a whole spectrum of reactions
in the cytosol.

We finally mention systems in which the terminal reaction event
requires the passage through intermediate target sites, as in the
transmission of signalling ions or proteins between neurons
\cite{1,2,3,sasha,sasha0,Reva}.  In the first step, calcium ions released on
the surface of the presynaptic bouton through voltage-activated
calcium channels diffuse towards specific sensor proteins to initiate
the release of neurotransmitters into the synaptic cleft.  In the
second step, the neurotransmitters diffuse towards receptors on the
boundary of the postsynaptic spine.  In the third step, a diffusive
signalling molecule is released in a bulbous head of the dendritic
spine and first needs to find an escape window, the first target site,
on the otherwise impermeable boundary.  This escape represents the
standard setup of the so-called Narrow Escape Problem
\cite{4,4a,4b,5,5a,5b,5c,6,6a,7,7a,7b,8,9,9a,9b,10,10a,100,100a,100b,100c,11,dist2}. Surmounting an entropy barrier \cite{11} the molecule
then enters into a narrow neck and thus may eventually reach the terminal target site,
the parent dendrite to which the neck is connected via its opposite extremity.

We here investigate the probability density function (PDF) of the time
instant at which the terminal reaction event in such a sequential
reaction cascade takes place.   In a basic setting, this random
time can be considered as the sum of {\it independent\/}
first-passage times (FPTs) associated with diffusion-reaction steps in
a relay race fashion.  As a consequence, its PDF is the
$N$-fold convolution of PDFs of each FPT, where $N$ is the number of
steps. Using Laplace transform techniques, one can access the PDF
through the inverse Laplace transform of the product of the generating
functions of individual FPTs. Even though the problem can be
considered as formally solved from the mathematical point of view,
most properties of such terminal FPTs necessitating a defined sequence
of preceding events are yet unknown, to large extent. Particularly, we
are interested in the shape of such FPT densities (FPTDs) in systems
of different dimensions.  What are their asymptotic short-time and
long-time behaviours? What are the relevant time scales of the
dynamics, and how do they depend on the system parameters? And,
importantly, how does the spatial arrangement of targets influence the
FPT PDF? Despite of the great biological relevance of such cascade
reactions,  the answers to these questions remain elusive.

We develop a mathematical framework for the general case with an \emph{arbitrary\/}
number of intermediate reactions taking place on sequentially labelled target sites
located at arbitrary fixed positions in space (figure \ref{fig:scheme}). Moreover we consider
both the case of perfect reactions of the messengers with their respective target
sites that happen with unit probability upon first mutual encounter, and the case
of imperfect reactions. For the latter, an elementary reaction act takes place with
a finite probability on each encounter, and thus in general necessitates repeated
reaction attempts interspersed with excursions away from the target. Concurrently
we consider here a somewhat simplified geometrical setup assuming that the cascade
of reactions takes place in an unbounded $d$-dimensional systems. Such a
simplification renders the derivations rather straightforward and the resulting
expressions are compact, permitting us to highlight the impact of the
intermediate messengers on the terminal reaction event in a transparent way, and
also to set an instructive framework for further analysis. Ultimately we obtain
explicit and rather simple formulae for the FPTD to the terminal point conditioned by
previous passages to an ordered set of positions in space at ordered time instants.
In some sense this result can be viewed as a kind of an analogue, formulated in terms of extreme events,   of the Wiener measure that defines the probability that the trajectory of a
Brownian motion visits some point at time instant $t$, given that earlier it visited
a fixed set of positions at fixed preceding times.  At the same time, it is worth stressing that  the
PDFs of the first-reaction events in unbounded systems do not posses integer moments of any order, while
they do in bounded domains,  and the latter define important characteristic
time-scales. However, in bounded domains the PDFs of the intermediate first-reaction
times  admit general spectral expansions whose forms depend on the relative distances from the
target sites to the domain boundaries (see, e.g., \cite{dist2,dist1,dist1a,dist4,dist5,dist3}).
This fact makes the analysis rather cumbersome and thus less transparent.
A discussion of the cascade reactions for  bounded geometries will be presented
elsewhere. 

This paper is outlined as follows. In section \ref{model} we formulate our model,
introduce basic notations and present general results. In section \ref{sec:1D} we
concentrate on one-dimensional systems with perfect or imperfect diffusion-controlled
reactions, while in section \ref{sec:2D} we present analogous results for
two-dimensional systems, followed by a discussion of the behaviour in three-dimensional
systems in section \ref{sec:3D} for which, in particular, we discuss the dependence of
the probability of incomplete reactions as a function of the number of
intermediate stages. Lastly in section \ref{conc} we conclude with a brief
summary of our results and outline some perspectives for further research.

\section{Model and general results} 
\label{model}

\begin{figure}
\centering
\includegraphics[width=7cm]{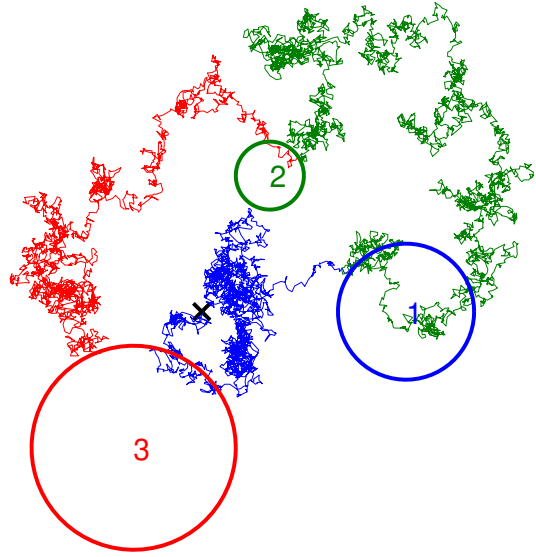}
\caption{
Illustration of the molecular relay race on the plane with three
circular \textit{perfect} immobile targets.  
The first particle starts from the
origin (black cross) and moves diffusively 
with the diffusion coefficient $D_1$
 (blue curve represents an individual simulated trajectory) until it hits the first
target (blue); then, the  second particle is launched and diffuses (green curve represents 
an individual simulated trajectory) with the diffusion coefficient $D_2$ 
until it finds 
 the second (green) target provoking a release of the third particle.  The relay race is terminated when the third particle, diffusing (see the red curve) with the diffusion coefficient $D_3$, arrives at the last target  (red). Note that for \textit{imperfect} reactions each particle may need to revisit the corresponding target  many times until a reaction event takes place. }
\label{fig:scheme}
\end{figure}

Consider a $d$-dimensional space, that is infinite in all directions,
with an arbitrary number $N$ of immobile target sites
(see figure \ref{fig:scheme} for an example with three target sites).  The targets are labelled sequentially by the
index $j=1,2,3,\ldots,N$. We refer to them as the first, the second,
the third target site, and so on. The target sites are assumed to be
spheres with the respective radii $r_j$, and $\mathbf{R}_j$ are the
vectors connecting the origin with their centres. The distance
$\rho_k$ between the centres of the $k$th and the $(k-1)$th target
sites is thus defined by $\rho_k=|\mathbf{R} _k-\mathbf{R}_{k-1}|$. We
use the convention that $\mathbf{R}_0=0$ (corresponding to the origin)
and hence, $\rho_1=|\mathbf{R}_1|$ is the distance from the origin to
the centre of the first target site.

Suppose next that  point-like particle (e.g., the first messenger)
launches from the origin at time instant $t_0=0$ and moves diffusively
with diffusion coefficient $D_1$ until it hits (the surface of) the
first target site. Then, with probability $p_1$, the first messenger
binds to this target site and triggers a release of the second
point-like particle (e.g., the second messenger) from the centre of
the first target site. At this moment, the first messenger is no
longer relevant for the further signal relay.\footnote{Note that some
signalling molecules may be inactivated after a certain lifetime
\cite{plos}, while others may have multiple functions, such as
"global" regulatory proteins \cite{mirny1}.}  If this reaction event
does not complete, with probability $1-p_1$, the first messenger keeps
on diffusing and hitting the surface of the first target site again
and again, until a successful binding event followed by the launch of
the second messenger.\footnote{In chemistry terms successful reactions
require the passage of an activation energetic barrier.}  We denote
this  random time instant as $\t_1$. The second messenger
diffuses with diffusion coefficient $D_2$ and seeks the second target
site. Once it arrives to this site, a binding event takes place with
probability $p_2$ and the third messenger is released. Similar to
before, the fate of the second particle is no longer of interest. In
the case of an unsuccessful reaction attempt, with probability $1-p_2$
the diffusive search process is repeated until a successful reaction
event takes place. The random duration of the second step is denoted
by $\t_2$. Such a cascade of diffusion-reaction processes proceeds
until the random moment in time when the $N$th particle successfully
binds to (with probability $p_N$) the surface of the $N$th target
site. We are interested here in the form of the probability density
function $H(t)$ of the total time $\T=\t_1+\t_2+\ldots+\t_N$, from the
initial release of the first messenger to the terminal reaction at the
$N$th target site: $\P\{\T\in(t,t+dt)\}=H(t)dt$.  For
simplicity, we will ignore the excluded volume of the targets, i.e.,
at each stage $j$, the particle can freely diffuse through all targets
$i \ne j$.  This simplifying assumption allows one to reduce the
$N$-target problem to a sequence of single target problems.

Let $\Psi_j(t)$, which is a function of the parameters $D_j$, $p_j$,
$r_j$ and $\rho_j$, denote the PDF of the event that it took the $j$th
particle exactly the time $t$ to bind to the $j$th target site and,
hence, to trigger the release of the $(j+1)$th particle, i.e.,
$\P\{\t_j\in(t,t+dt)\}=\Psi_j(t)dt$.  Correspondingly, let
$\tilde{\Psi}_j(p)=\mathscr{L}\left\{\Psi_j(t)\right\} = \int^{\infty}_0 dt \, e^{-pt}\, \Psi_j(t)$ denote
its Laplace transform, i.e., the generating function of the FPT
$\t_j$.  As the $\t_j$ are considered as independent random
variables, then, quite formally, one can write down the desired PDF
$H(t)$ as a convolution
\begin{equation}
\label{a}
H(t)=\int^t_0dt_{N-1}\int^{t_{N-1}}_0dt_{N-2}\ldots\int^{t_2}_0dt_1\prod_{j=1}^N
\Psi_j(t_j-t_{j-1}),\quad t_N=t,\,\,\, t_0=0, 
\end{equation}
which implies that the generating function of $\T$ is simply the product of the
generating functions of $\t_j$,
\begin{equation}
\label{b}
\tilde{H}(p)=\prod_{j=1}^N\tilde{\Psi}_j(p).
\end{equation}
Several remarks at this point are in order:\\ (i) Expressions \eqref{a} and
\eqref{b} are fairly general and valid not only for standard diffusive motion
but also for anomalous diffusion. \\(ii) A natural reset of memory that happens
with reactions occurring at intermediate steps within the cascade makes this
model applicable to non-Markovian processes with a memory, such as, e.g., a
fractional Brownian motion.\\ (iii)  In some settings, a release of the $(j+1)$th particle upon
arrival of the $j$th messenger to the corresponding target site does not occur
instantaneously but may require some random time $\delta_j$ (with PDF $\psi_j(
t)$) to produce the corresponding effector and to eventually synthesise the
$(j+1)$th particle. In this case expression \eqref{a} is to be appropriately
generalised to include such intermediate stages but will still maintain the
form of a convolution. As a result expression \eqref{b} is to be multiplied by
the product $\prod_{j=2}^{N-1}\tilde{\psi}_j(p)$, in which $\tilde{\psi}_j(p)$
denotes the generating function of $\delta_j$. We note, however, that these
purely chemical processes usually do not involve a diffusive transport stage
and hence happen at much shorter time scales, which can thus be safely discarded.\\
(iv) Expression \eqref{b} implies a simple relation between the $n$th
order moment of the total time $\T$ and the moments of the
first-reaction times $\t_j$ of the intermediate stages within a
cascade, \textit{if all of them exist}, as it happens, e.g., for
search processes in bounded domains. This follows straightforwardly 
from the independence of the first-reaction times of the intermediate stages and can be derived, e.g., by simply
differentiating $n$ times both sides of this expression with respect
to $p$ and ultimately setting $p =0$. Quite trivially, one has
\begin{equation}
\label{1st}
\overline{\T}=\overline{\t_1}+\overline{\t_2} +\ldots+\overline{\t_N}  \,,
\end{equation}
where the overbar denotes averaging. Differentiating further, one
finds that also the variance of $\T$ is just the sum of variances of
the first-reaction times $\t_j$ and moreover, a cumulant of an
arbitrary order of the PDF of $\T$ is the sum of cumulants of $\t_j$
of the same order. Note, however, that even in bounded domains the
moments of first-passage or reaction times are often not
representative of actual behaviour \cite{mat1,mat2} and stem from
anomalously long searching trajectories, corresponding to long-$t$
tails of the PDF. As a consequence, the moments or cumulants alone do
not tell us much, if anything about a typical behaviour.  We will
return to this question at the end of this paper. \\
(v) There are some subtleties of the behaviour in high-dimensional
infinite systems ($d \geq 3$), in which the fractal dimension of
trajectories (equal to $2$ in the case of standard Brownian motion) is
smaller than the embedding spatial dimension. In such systems, at each
stage there is a finite probability that the reaction does not take
place at all because a finite fraction of trajectories travels to
infinity and never reaches the corresponding target site.  We
will address this question in section \ref{sec:3D}, while sections
\ref{sec:1D} and
\ref{sec:2D} focus on the form of the PDF for one- and two-dimensional systems, 
in which the reaction events happen with probability $1$.

In the following, we consider the general case of imperfect reactions such that
all (or some of) reaction probabilities $p_j$ are less than unity, and thus the
particle may have repeated collisions with the target until a successful reaction
occurs. The calculation of the corresponding probability density function of the first
reaction time $\T$ thus amounts to solving the diffusion equation with radiation
boundary condition (also called Robin or mixed boundary condition) on the surface
of the target site, see, e.g., \cite{Collins49,denis,mihail}. Here the diffusive current
through the surface is balanced by the probability density right at the surface,
the proportionality factor being the reactivity $\kappa_j$. This reactivity is
related to the reaction probability $p_j$ via $\kappa_j=p_jv/(1-p_j)$, where $v$
has a dimension of velocity and equals the thickness of the reaction zone around
the target site multiplied by the frequency of reaction attempts upon a contact
with the surface, see \cite{redner,burl,benichou,Lawley19,Grebenkov20} for more details. The reactivity is
infinite when the corresponding reaction probability is unity, $p_j=1$. In turn,
the reactivity equals zero when $p_j=0$, i.e., the reaction is completely inhibited.
As the cascade reaction never occurs if at least one of the $p_j$ is zero, we assume
that all $p_j$ (and thus all $\kappa_j$) are strictly positive: $\kappa_j>0$. We
also assume that all diffusion coefficients are finite and strictly positive: $0<
D_j<\infty$. While targets can be point-like in a one-dimensional setting, one has
to consider extended targets in higher dimensions so that $r_j>0$. We also assume
that $\rho_j>r_j$.

\section{One-dimensional systems}
\label{sec:1D}

We start with one-dimensional systems, in which  two situations
have to be distinguished: (i) in the so-called one-sided case, a
particle arriving at the target site is reflected when a reaction
attempt is rejected (with probability $1-p_j$) in the direction from
which it arrived to this site -- in other words, a particle cannot pass
through the target site; (ii) in the two-sided case, when the
reaction at the target site does not occur, a particle passes through
this site and may approach the same target from the opposite direction. In
the one-sided case the generating function and the PDF corresponding
to the $j$th stage of the cascade reaction process are given by (see,
e.g., \cite{dist4})
\begin{flalign}
\label{im1}
\tilde{\Psi}_j(p)&=\left(1+\frac{\sqrt{D_jp}}{\kappa_j}\right)^{-1}\exp\left(-
\rho_j\sqrt{p/D_j}\right),\\
\Psi_j(t)&=\frac{\kappa_j}{\sqrt{\pi D_jt}}\exp\left(-\frac{\rho_j^2}{4D_jt}\right) 
\left(1-\kappa_j\sqrt{\frac{\pi t}{D_j}}\erfcx\left(\frac{\rho_j}{2\sqrt{D_jt}}+
\kappa_j \sqrt{\frac{t}{D_j}}\right)\right), 
\end{flalign}
where $\erfcx(z)=e^{z^2}\erfc(z)$ is the scaled complementary error
function, while $\erfc(z)$ is the complementary error function \cite{abr}.

In order to determine explicitly the PDF of the time of the terminal
reaction event in a cascade of imperfect reactions in one-dimensional
systems, we first assume that all ratios $\sqrt{D_j}/\kappa_j$ are
different from one another (the case when all of them are equal will
be considered below). In this case it is convenient to expand the
generating function into partial fractions,
\begin{flalign}
\label{s}
\tilde{H}(p)&=\left(\prod_{j=1}^N\left(1+\dfrac{\sqrt{D_jp}}{\kappa_j}\right)
\right)^{-1}\exp\left(-\sqrt{Tp}\right)=\exp\left(-\sqrt{Tp}\right)\sum_{j=1}^N
\dfrac{\pi_j}{1+\sqrt{D_jp}/\kappa_j}, 
\end{flalign} 
with
\begin{equation}
\label{eq:pij}
\pi_j=\prod_{i=1,i\neq j}^N\left(1-\frac{\kappa_j}{\kappa_i}\sqrt{\frac{D_i}{D_j}}
\right)^{-1},
\end{equation}
and the naturally emerging characteristic time scale
\begin{equation}
\label{time}
T=\left(\sum_{j=1}^N\frac{\rho_j}{\sqrt{D_j}}\right)^2.
\end{equation}
This characteristic time tends to infinity when any of
the diffusion coefficients vanishes (or  any of the $\rho_j$ tends to
infinity); in turn, the contribution of the $j$th stage vanishes when
either $\rho_j=0$ (the target sites overlap) or $D_j=\infty$, as it
should be.

The Laplace transform can now be readily inverted to give
\begin{flalign}
\label{full1d}
H(t)&=\frac{1}{\sqrt{\pi t}}\exp\left(-\frac{T}{4t}\right)\sum_{j=1}^N\frac{\kappa_j
\pi_j}{\sqrt{D_j}}\left(1-\kappa_j\sqrt{\frac{\pi t}{D_j}}\erfcx\left(\frac{1}{2}
\sqrt{\frac{T}{t}}+\kappa_j\sqrt{\frac{t}{D_j}}\right)\right).
\end{flalign}
This is an exact expression which holds for arbitrary $t$ and
arbitrary sets $\{D_j\}$ and $\{\kappa_j\}$ (or equivalently
$\{p_j\}$), conditioned by the constraint that all ratios
$\sqrt{D_j}/\kappa_j$ are unequal to each other.  Albeit equation
\eqref{full1d} defines the PDF explicitly, it contains special
functions, and it might be useful to make the behaviour more explicit
in the limits $t\to\infty$ and $t\to0$. In the limit $t\to\infty$, the
PDF behaves to leading order as
\begin{align}
\label{tail}
H(t)\sim
\dfrac{\sum_{j=1}^N \biggl(\dfrac{\rho_j}{\sqrt{D_j}}
+\dfrac{\sqrt{D_j}}{\kappa_j}\biggr)}{\sqrt{4\pi t^3}} \qquad (t\to\infty),
\end{align}
where we have used the identities $\sum_{j=1}^N\pi_j=1$ and $\sum_{j=1}^N\sqrt{D_j}
\pi_j/\kappa_j=\sum_{j=1}^N\sqrt{D_j}/\kappa_j$. Note that we use here and henceforth the symbol $\sim$
for asymptotic equivalence. 
We emphasise two consequences:
(i) It is well  known (see, e.g., \cite{redner,burl}) that for
diffusion-controlled reactions in low-dimensional systems   the
asymptotic behaviour of the apparent reaction constants,  calculated within a suitably generalised Smoluchowski approach,  is independent
of the intrinsic chemical rate $\kappa$. This is a direct consequence
of the fact that in such systems Brownian motion visits each point in
the system many times, i.e., oversamples the space. Hence, as time
evolves it becomes progressively more difficult to transport a
particle diffusively on larger and larger spatial scales.\footnote{In
facilitated diffusion this fact is circumvented by intermittent bulk
excursions \cite{bvh}, while in L{\'e}vy flight search similar
oversampling is prevented due to long-tailed jump length PDFs leading
to leapovers across a target \cite{koren,vladnjp}.}  As a result the
diffusive delivery of a particle to the target site poses an
increasingly more dominant resistance to the reaction process than the
resistance due to the reaction activation barriers embodied in the
$\kappa_j$. In the first-passage time formulation, however, this
appears not to be the case --  even the long-time behaviour of the PDF
of the first reaction event time contains an explicit dependence on
all $\kappa_j$. (ii) For fixed $\rho_j$ and $\kappa_j$, the prefactor
in equation \eqref{tail} is a non-monotonic function of $D_j$, which
diverges when either $D_j\to\infty$ or $D_j\to0$. The prefactor
attains a minimal value when $D_j=D^*_j=\kappa_j\rho_j$. (iii)
Finally, the $H(t)\simeq t^{-3/2}$ scaling is the same as for the
L{\'e}vy-Smirnov density for first-passage in a semi-infinite domain
\cite{redner}, independent of $N$.
 
In the short-time limit, one can expand the exact expression
\eqref{full1d} at small $t$ and take into account a rather interesting
property of $\pi_j$: namely, the sums
$\sum_{j=1}^N\pi_j(\sqrt{D_j}/\kappa_j)^n=0$ for any positive integer
$n<N$, and
$\sum_{j=1}^N\pi_j(\sqrt{D_j}/\kappa_j)^N=(-1)^{N+1}\prod_{j=1}^N\sqrt{D_j}/
\kappa_j$.  The PDF \eqref{full1d} to leading order behaves as
\begin{equation}
\label{shorta}
H(t)\sim\frac{2^{N-1}}{\sqrt{\pi}} \frac{t^{N-3/2}}{T_{\kappa}^{N/2} \, T^{(N-1)/2}}\exp\left(-\frac{T}{4t}\right),\quad
t\ll\min\left(D_j/\kappa_j^2\right),
\end{equation}
where 
\begin{equation}  \label{eq:Tkappa}
T_\kappa = \bigl(D_1\cdots D_N/(\kappa_1^2\cdots \kappa_N^2)\bigr)^{1/N}
\end{equation}
is another time scale, which is the geometric mean of times
$D_j/\kappa_j^2$.  While $T$ characterised exclusively the diffusive
steps, $T_\kappa$ accounts for the interplay between diffusion and
reaction on each target.  Here we see that the exponential cutoff
$\exp(-T/[4t])$, similar to the L{\'e}vy-Smirnov form, is modulated by
the $N$-dependent power $t^{N-3/2}$.

Lastly, we consider the particular case when all $D_j=D$ and
$\kappa_j=\kappa$  -- the case in which the ratio in the first line in
equation \eqref{s} cannot be decomposed into elementary fractions. In
this special case expression \eqref{s} attains the form
\begin{align}
\tilde{H}(p)&=\left(1+\frac{\sqrt{Dp}}{\kappa}\right)^{-N}\exp\left(-\Rho\sqrt{p/D}
\right),
\end{align}
where $\Rho=\sum\nolimits_j\rho_j$. Inverting the latter expression we find that
here the PDF is given by
\begin{align}
\label{par}
H(t)&=\frac{1}{\sqrt{4\pi t^3}}\left(\frac{2\kappa^2t}{D}\right)^{N/2}\exp\left(-
\frac{\Rho^2}{4Dt}+\frac{\left(2\kappa t+\Rho\right)^2}{8Dt}\right)\nonumber\\
&\times\Bigg(\frac{\Rho}{\sqrt{D}}{\rm D}_{-N}\left(\frac{2\kappa t+\Rho}{\sqrt{2
Dt}}\right)+N\sqrt{2t} \, \mathrm{D}_{-N-1}\left(\frac{2\kappa t+\Rho}{\sqrt{2Dt}}\right)
\Bigg),
\end{align}
where $\mathrm{D}_{-\alpha}(z)$ is the parabolic cylinder function
\cite{abr}. Note that in both limits $t\to0$ and $t\to\infty$ the
argument of the parabolic cylinder function becomes infinitely
large. The asymptotic behaviour of $\mathrm{D}_{-\alpha}(z)$ in the
large-$z$ limit to leading order follows \cite{abr}
\begin{equation}
\mathrm{D}_{-\alpha}(z)\sim z^{-\alpha}\exp\left(-\frac{z^2}{4}\right)\quad (z\to
\infty),
\end{equation}
such that $H(t)$ attains the asymptotic form
\begin{equation}
H(t)\sim\frac{2^{N-1}\kappa^Nt^{N-3/2}}{\sqrt{\pi D}\left(2\kappa t+\Rho\right)^N} 
\left(\Rho+\frac{2NDt}{2\kappa t+\Rho}\right)\exp\left(-\frac{\Rho^2}{4
Dt}\right),
\end{equation}
which is valid for both $t\to\infty$ and $t\to0$. Keeping only the leading terms we
get
\begin{equation}
H(t)\sim\frac{2^{N-1}\kappa^Nt^{N-3/2}}{\sqrt{\pi D}\Rho^{N-1}}\exp\left(-\frac{
\Rho^2}{4Dt}\right)\qquad (t\to 0)
\end{equation}
and
\begin{equation}
H(t)\sim\frac{\left(\Rho+ND/\kappa\right)}{\sqrt{4\pi Dt^3}} \qquad (t\to\infty).
\end{equation}
Thus, despite the fact that the distributions \eqref{full1d} and \eqref{par}
pertain to different physical situations and thus have different functional
forms, the one in equation \eqref{full1d} defines exactly the same asymptotic
behaviour in the limits $t\to0$ and $t\to\infty$ as expression \eqref{par}. This
can be readily seen by setting $D_j=D$ and $\kappa_j=\kappa$ in equations
\eqref{tail} and \eqref{shorta}.

We now turn to the two-sided case with imperfect reactions, in which a
particle can pass through a target. The expression analogous to
expression \eqref{s} then reads
\begin{flalign}
\label{ss}
\tilde{H}(p)&=\left(\prod_{j=1}^N\left(1+2\dfrac{\sqrt{D_jp}}{\kappa_j}\right)
\right)^{-1}\exp\left(-\sqrt{Tp}\right).
\end{flalign}
Inspecting the latter formula and comparing it against expression
\eqref{s}, we conclude that all the above results obtained for the
one-sided case are valid upon the simple replacement $\kappa_j \to
\kappa_j/2$ , i.e., each intrinsic reactivity just gets
reduced by a factor $1/2$. At first glance, such a reduction of the
reactivity may seem to be somewhat counter-intuitive.  Indeed,
permitting for the reaction to happen from both sides of the target
site, one effectively increases the size of the target which, in
principle, should ease reaction events. A plausible explanation is
that permitting the particle to pass through the target also gives it
the  possibility to escape from it to both plus and minus infinity, while
in the one-sided case it can go to infinity only in one direction.

Finally, the above expressions are considerably simplified in the ideal case of
perfect reactions when a successful reaction event, resulting in the release of
the subsequent messenger, takes place instantaneously upon the first encounter
with the surface, $p_j=1$. Hence all $\t_j$ are merely the first-passage times
from ${\bf R}_{j-1}$ to ${\bf R}_j$. In this case we obtain
\begin{equation}
\label{term1d}
\tilde{H}(p)=\exp\left(-\sqrt{Tp}\right), \qquad H(t)=\sqrt{\frac{T}{4\pi t^3}}\exp
\left(-\frac{T}{4 t}\right),
\end{equation}
with $T$ given by equation \eqref{time}. These expressions can be either deduced
from expression \eqref{im1} in the limit $\kappa_j\to\infty$ or found directly
from the L\'evy-Smirnov PDF and generating function of the first-passage time
$\t_j$ to the perfect target,
\begin{equation}
\label{ls}
\tilde{\Psi}_j(p)=\exp\left(-\rho_j\sqrt{p/D_j}\right), \qquad \Psi_j(t)=\frac{
\rho_j}{\sqrt{4\pi D_jt^3}}\exp\left(-\frac{\rho_j^2}{4D_jt}\right). 
\end{equation}
One can see that the PDF $H(t)$ of the terminal reaction event preserves the form
of the individual components, corresponding to the intermediate stages in reaction
cascade. In fact expression \eqref{term1d} has exactly the L\'evy-Smirnov form in
which the characteristic time $T$ includes the parameters representing the
intermediate stages.

\section{Two-dimensional systems} 
\label{sec:2D}

The generating function of the first-passage time to a partially reactive circular
target in a plane is (see, e.g., \cite{dist4})
\begin{equation}
\label{int}
\tilde{\Psi}_j(p)=\frac{K_0\left(\rho_j\sqrt{p/D_j}\right)}{K_0\left(r_j\sqrt{p/D_j}
\right)+\dfrac{\sqrt{pD_j}}{\kappa_j}K_1\left(r_j\sqrt{p/D_j}\right)},
\end{equation}
where $K_\nu(z)$ is the modified Bessel function of the second kind
\cite{abr}.\footnote{One immediately sees from $K_0(0)=\infty$ that the
first-passage to a point-like target in two dimensions is impossible.}
Even though the inverse Laplace transform of expression \eqref{int} formally
yields the PDF,
\begin{equation}
\Psi_j(t)=\L^{-1}\left\{\tilde{\Psi}_j(p)\right\},
\end{equation}
this inversion cannot be performed analytically and cannot even be written in
terms of common special functions. One thus has to resort to numerical analysis
or concentrate on the asymptotic behaviour of the PDF in the limits $t\to0$ and
$t\to\infty$ (see, e.g., \cite{Levitz08,dist4}). For these purposes it is convenient to represent the inverse Laplace
transform as the Bromwich integral and to exploit the regularity property of the
$K_\nu(z)$ in order to change the integration variable and to deform the integration
contour in the complex plane. These operations allow one to express $\Psi_j(t)$ as
the integral over the positive real half-axis (see \cite{Grebenkov21} for details),
\begin{equation}
\label{brom}
\Psi_j(t)=-\frac{2}{\pi t}\int\limits_0^\infty dz \, z \, e^{-z^2}\Im\left(\dfrac{A\left(
zr_j/\sqrt{D_jt},\rho_j/r_j\right)}{1+\dfrac{D_j}{\kappa_jr_j}B\left(r_jz/\sqrt{D_j
t}\right)}\right).
\end{equation}
Here ${\rm Im}$ denotes the imaginary part of the expression and
\begin{eqnarray}
A(z,r)&=&\frac{(J_0(z)+i Y_0(z))(J_0(zr)-iY_0(zr))}{J_0^2(z)+Y_0^2(z)},\\
B(z)&=&\frac{z(J_0(z)J_1(z)+Y_0(z)Y_1(z))+i\frac{2}{\pi}}{J_0^2(z)+Y_0^2(z)},
\end{eqnarray} 
where $J_\nu(z)$ and $Y_\nu(z)$ are the Bessel functions of the first and second
kind, respectively \cite{abr}. Following \cite{Grebenkov21} we next generalise
expression \eqref{brom} for the reaction cascade case. According to expression
\eqref{b} the generating function of the first-passage time to the terminal reaction event 
in a cascade reads
\begin{equation}
\label{eq:Hp}
\tilde{H}(p)=\prod\limits_{j=1}^N\dfrac{K_0\left(\rho_j\sqrt{p/D_j}\right)}{K_0
\left(r_j \sqrt{p/D_j}\right)+\dfrac{\sqrt{pD_j}}{\kappa_j}K_1\left(r_j\sqrt{p/
D_j}\right)}.
\end{equation}
Since the product of ratios of modified Bessel functions of the second kind has
no poles one can apply the same technique as above to get the following result:
\begin{equation}
\label{eq:Ht_2D}
H(t)=-\frac{2}{\pi t}\int\limits_0^\infty dz\,z \, e^{-z^2}\Im\left(\prod\limits_{j=1}^N 
\dfrac{A\left(zr_j/\sqrt{D_jt},\rho_j/r_j\right)}{1+\dfrac{D_j}{\kappa_jr_j}B\left(
r_jz/\sqrt{D_jt}\right)}\right),
\end{equation}
Equation \eqref{eq:Ht_2D} is exact for any $t$ and any values of the
parameters, and generalises the Bromwhich-type representation of the
PDF of first-reaction times for the problem with a single imperfect
target site (see, e.g., \cite{dist4}) to the case of a cascade of
diffusion controlled reactions with an array of targets.

The short-time behaviour of the PDF \eqref{eq:Ht_2D} can be obtained by analysing
the large-$p$ limit of the generating function $\tilde{H}(p)$ in expression
\eqref{eq:Hp}. In this limit, $K_\nu(z)\sim e^{-z}\sqrt{\pi/(2z)}$ \cite{abr} and
hence
\begin{align}
\tilde{H}(p)\sim\dfrac{\left(\prod_{j=1}^N\dfrac{r_j}{\rho_j}\right)^{1/2}\left(
\prod_j'\dfrac{\kappa_j}{\sqrt{D_j}}\right)}{p^{N'/2}}\exp\left(-\sqrt{pT}\right),
\end{align}
where 
\begin{equation}
\label{eq:T_2d}
T=\left(\sum\limits_{j=1}^N\frac{\rho_j-r_j}{\sqrt{D_j}}\right)^2 
\end{equation}
is a characteristic time scale which only slightly differs from our result
\eqref{time} above, $N'$ is the number of finite $\kappa_j$, and $\prod
\nolimits_j'$ denotes the product extending over such values of $j$ for
which the corresponding $\kappa_j$ is finite. The inverse Laplace transform
yields the short-time asymptotic form
\begin{align}
\label{eq:Ht_short}
H(t)\sim\left(\dfrac{4}{T}\right)^{(N'-1)/2}\left(\prod_{j=1}^N\frac{r_j}{\rho_j}
\right)^{1/2}\left(\prod\nolimits_j'\frac{\kappa_j}{\sqrt{D_j}}\right)\frac{t^{N'
-3/2}}{\sqrt{\pi}}\exp\left(-\frac{T}{4t}\right).
\end{align}
When all elementary reaction steps are imperfect, that is, $\kappa_j<\infty$ for
all $j$, one has $N'=N$ and the corresponding short-time behaviour follows from
equation \eqref{eq:Ht_short} by replacing $N'$ by $N$. In the opposite situation
when all reaction steps are perfect, i.e., $\kappa_j=\infty$ for all $j$, $N'=0$
and, hence, the product $\prod\nolimits_j'\left(\kappa_j/\sqrt{D_j}\right)=1$
which yields
\begin{equation}
\label{eq:Ht_short_perfect}
H(t)\sim\frac{\left(\prod_{j=1}^N\left(r_j/\rho_j\right)\right)^{1/2}\sqrt{T}}{
\sqrt{4\pi t^3}}\exp\left(-\frac{T}{4t}\right) \quad (t\to0).
\end{equation}
Therefore, the short-time tail of $H(t)$ in two-dimensional systems
with perfect reactions has the form of the L\'evy-Smirnov density,
analogous to one-dimensional systems, and differs from that case only by the
dependence of the prefactor and of the characteristic time on system
parameters.

\begin{figure}
\centering
\includegraphics[width=75mm]{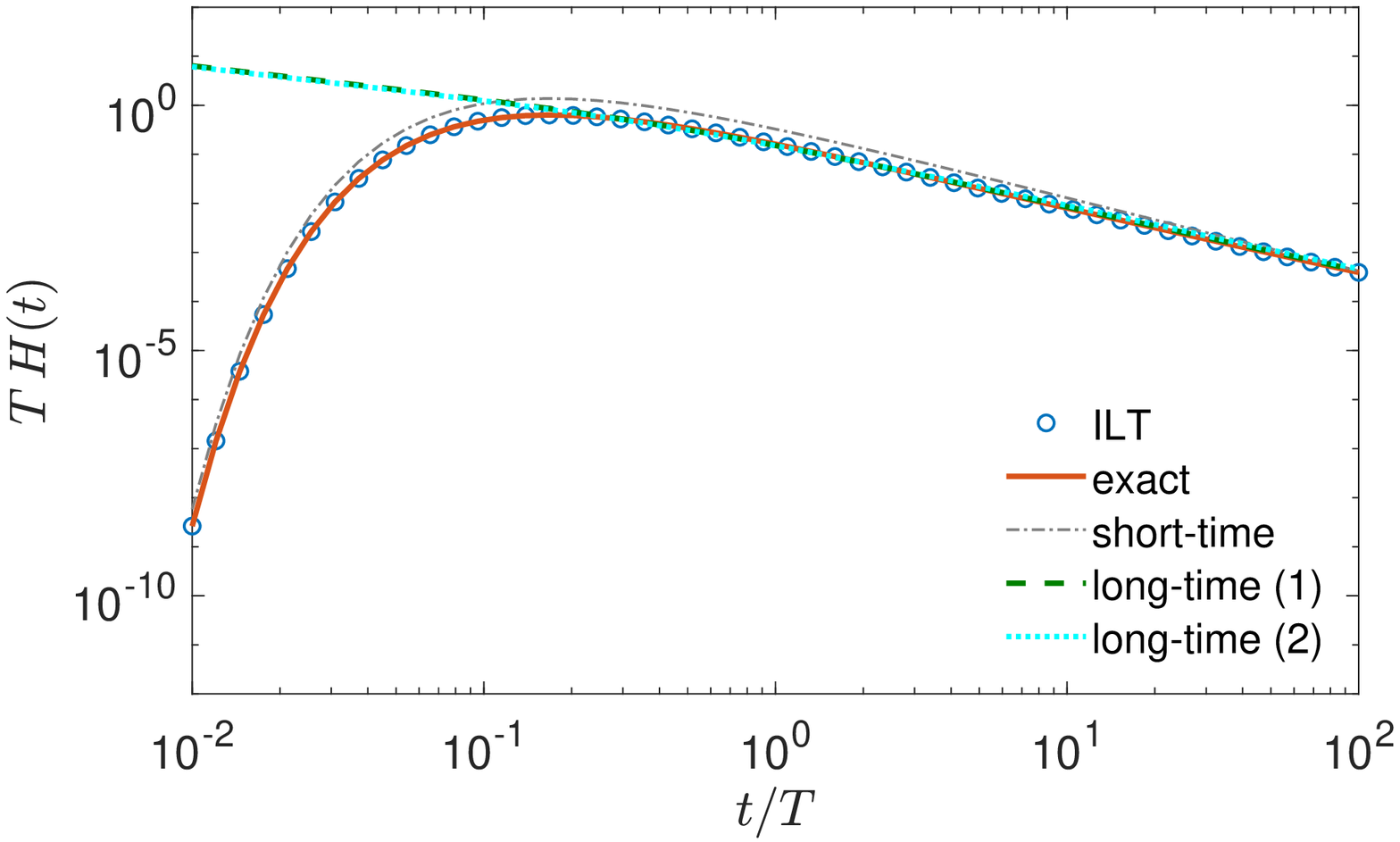}
\includegraphics[width=75mm]{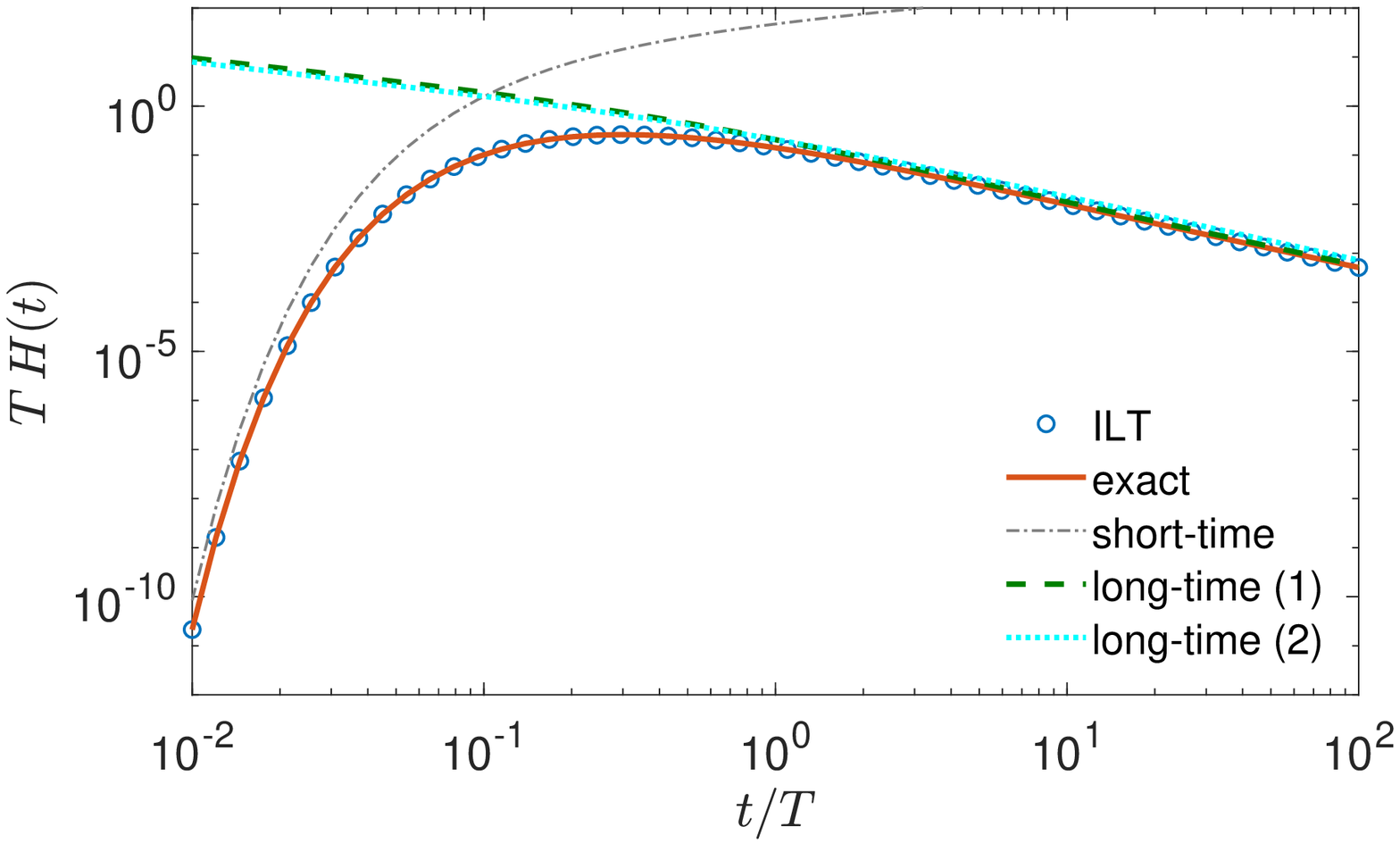}
\caption{ Rescaled 
PDF $T H(t)$ as function of dimensionless time $t/T$ for a 2-stage cascade ($N=2$) in two dimensions with
$r_1=1$, $\rho_1=2$, $D_1=1$, and $r_2=1.8$, $\rho_2=2$,
$D_2=0.4$. {\bf Left:} $\kappa_1=\kappa_2=\infty$ (perfect reactions); {\bf Right:}
$\kappa_1=10$, $\kappa_2=1$  (imperfect reactions). Circles represent the numerical Laplace
inversion (abbreviated as ILT) of equation \eqref{eq:Hp} with $N=2$ using the Talbot
algorithm; solid lines show our exact representation \eqref{eq:Ht_2D};
dash-dotted lines represent the short-time asymptote
\eqref{eq:Ht_short_perfect} for perfect reactions, and
\eqref{eq:Ht_short} for imperfect reactions, whereas dashed
lines indicate the long-time asymptotic relations
\eqref{eq:Ht_2D_long} and \eqref{eq:Ht_2D_long2}.   }
\label{fig:Ht_2D_N2} 
\end{figure}

The analysis of the long-time behaviour of $H(t)$ is generally much more subtle
(see, e.g., \cite{Grebenkov21}) and here the general integral representation
\eqref{eq:Ht_2D} appears to be particularly useful. Following \cite{Grebenkov21}
we notice that the long-time limit corresponds to $z\to0$, for which
\begin{equation}
A(z,r)\sim1-B(z)\ln r,\qquad B(z)\sim\frac{1}{\ln(1/z)+\ln2-\gamma-\pi i/2} \,,
\end{equation}
where $\gamma$ is the Euler-Mascheroni constant. Substituting these
expressions into equation \eqref{eq:Ht_2D} and taking advantage of
some standard arguments (see \cite{Grebenkov21}) we find the following
long-$t$ asymptotic
\begin{equation}
\label{eq:Ht_2D_long}
H(t)\sim-\frac{1}{\pi t}\Im\left(\prod\limits_{j=1}^N\dfrac{1-B_j(t)\ln(\rho_j
/r_j)}{1+\dfrac{D_j}{\kappa_jr_j}B_j(t)}\right),
\end{equation}
where
\begin{equation}
B_j(t)=\frac{1}{\ln(\sqrt{D_jt}/r_j)+\ln2-\gamma-\pi i/2}.
\end{equation}
We note that the asymptotic form \eqref{eq:Ht_2D_long} turns out to be remarkably
accurate for sufficiently large $t$, as evidenced by comparison with the numerical
inversion of the Laplace transform of expression \eqref{eq:Hp}, see figure
\ref{fig:Ht_2D_N2}. Expression \eqref{eq:Ht_2D_long} contains, however, an imaginary
part of some rational function, which somewhat obscures its actual time dependence,
and it can thus be instructive to calculate this imaginary part explicitly. To this
end, we rewrite $B_j(t)$ formally as
\begin{equation}
B_j(t)=2\frac{\ln(4e^{-2\gamma}D_jt/r_j^2)+\pi i}{\ln^2(4e^{-2\gamma}D_jt/r_j^2)+
\pi^2},
\end{equation}
which implies that $B_j(t)$ tends to zero as $t\to\infty$. Expanding then the
product in equation \eqref{eq:Ht_2D_long} in inverse powers of this small
parameter and keeping only the leading-order terms we find
\begin{equation} \label{eq:Ht_2D_long2}
H(t)\sim\frac{2}{t}\sum\limits_{j=1}^N\dfrac{\dfrac{D_j}{\kappa_jr_j}+\ln(\rho_j
/r_j)}{\ln^2(4e^{-2\gamma}D_jt/r_j^2)+\pi^2}\qquad (t\to\infty),
\end{equation}
which defines the long-time limit explicitly. This asymptotic form is valid for
arbitrary values of $\{\rho_j\}$, $\{r_j\}$, $\{\kappa_j\}$, and $\{D_j\}$. For
perfect reactions, when all $\kappa_j=\infty$ (and $D_j>0$), the first term in
the numerator vanishes such that the summands becomes proportional to the
logarithm of the ratio $\rho_j/r_j$ and logarithmically weakly depend on the
diffusion coefficients.

\section{Three-dimensional systems}
\label{sec:3D}

In three dimensions the generating function of $\t_j$ reads
\begin{equation}
\label{eq:Psi_3d}
\tilde{\Psi}_j(p)=\frac{r_j}{\rho_j}\left(1+\dfrac{D_j}{\kappa_jr_j}+\dfrac{1}{
\kappa_j}\sqrt{pD_j}\right)^{-1}e^{-(\rho_j-r_j)\sqrt{p/D_j}},
\end{equation}
and the PDF has the closed form
\begin{align}  
\Psi_j(t)&=\frac{\kappa_j}{\rho_j}e^{-(\rho_j-r_j)^2/(4D_jt)}\biggl\{\frac{r_j}{
\sqrt{\pi D_jt}}\nonumber\\
&-(1+\kappa_jr_j/D_j)\erfcx\biggl(\frac{\rho_j-r_j}{\sqrt{4D_jt}}+(1+\kappa_jr_j/
D_j)\frac{\sqrt{D_jt}}{r_j}\biggr)\biggr\}.
\end{align}
Rewriting expression \eqref{eq:Psi_3d} formally as
\begin{equation}
\label{eq:Psi_3d_bis}
\tilde{\Psi}_j(p)=\dfrac{r_j}{\rho_j\left(1+\dfrac{D_j}{\kappa_jr_j}\right)}\left(
1+\dfrac{1}{\kappa_j+D_j/r_j}\sqrt{pD_j}\right)^{-1}e^{-(\rho_j-r_j)\sqrt{p/D_j}},
\end{equation}
we observe that result \eqref{eq:Psi_3d_bis} indeed has exactly the
same functional form as $\tilde{\Psi}_j(p)$ in the one-dimensional
case (see equation \eqref{im1}) if one replaces $\rho_j$ by
$\rho'_j=\rho_j-r_j$, $\kappa_j$ by $\kappa'_j=\kappa_j +D_j/r_j$, and
multiplies  expression \eqref{im1} by $(r_j/\rho_j)
(1+D_j/(\kappa_j r_j))^{-1}$. As a consequence, we can use the results
from section \ref{sec:1D} for the PDF $H(t)$ and its asymptotic
behaviour in the three-dimensional case. For instance, once all
$\sqrt{D_j}/\kappa'_j$ are distinct, equation \eqref{full1d} in the
three-dimensional case has the form
\begin{flalign}
\label{full_3d}
H(t)&=P_{\mathrm{react}}\frac{\exp(-T/[4t])}{\sqrt{\pi t}}\sum_{j=1}^N\frac{\kappa'
_j\pi_j}{\sqrt{D_j}}\left(1-\kappa'_j\sqrt{\frac{\pi t}{D_j}}\erfcx\left(\frac{1}{2}
\sqrt{\frac{T}{t}}+\kappa'_j\sqrt{\frac{t}{D_j}}\right)\right),
\end{flalign}
in which 
\begin{equation}
\label{react}
P_{\mathrm{react}}=\prod\limits_{j=1}^N\dfrac{r_j}{\rho_j\left(1+\dfrac{D_j}{\kappa_j
r_j}\right)},
\end{equation}
and the time scale $T$ is given by equation \eqref{eq:T_2d}. Moreover
$\pi_j$ is determined by equation \eqref{eq:pij} with $\kappa_j$
replaced by $\kappa'_j$.  Finally, one retrieves the same short-time
and long-time asymptotic behaviour of $H(t)$ as in the one-dimensional
case, where the prefactor is corrected by $P_{\mathrm{react}}$, 
\begin{equation}
\label{eq:Ht_3D_short}
H(t)\sim\frac{2^{N-1}}{\sqrt{\pi}} \frac{t^{N-3/2}}{T_{\kappa}^{N/2} \, T^{(N-1)/2}}\exp\left(-\frac{T}{4t}\right) \qquad
(t\to 0)
\end{equation}
and
\begin{align}
\label{eq:Ht_3D_long}
H(t)\sim P_{\mathrm{react}} \dfrac{\sum_{j=1}^N \biggl(\dfrac{\rho_j}{\sqrt{D_j}}
+\dfrac{\sqrt{D_j}}{\kappa'_j}\biggr)}{\sqrt{4\pi t^3}} \qquad (t\to\infty),
\end{align}
where $T_\kappa$ is given by \eqref{eq:Tkappa} with $\kappa_j$
replaced by $\kappa'_j$.

Despite of these similarities the three-dimensional behaviour is crucially different
due to the fact that the diffusing messenger can escape to infinity at each
diffusion-reaction stage of a cascade. Indeed, setting $p\to0$ in expression
\eqref{eq:Psi_3d} one gets the reaction probability $\tilde{\Psi}_j(0)=(r_j/
\rho_j)/(1 + D_j/(\kappa_jr_j))$, which is strictly smaller than $1$ as soon
as $r_j<\rho_j$. In other words, the PDF $\Psi_j(t)$ is not normalised to
unity. As a consequence, the PDF $H(t)$ is also not normalised,
\begin{equation}
\int\limits_0^\infty dt \,H(t)=\tilde{H}(0)=P_{\mathrm{react}}<1.
\end{equation}
Therefore, $P_{\mathrm{react}}$ in equation \eqref{react} is an important 
characteristic property which defines, for an arbitrary set of system parameters, 
the fraction of trajectories in the cascade that eventually arrive to the terminal
target site, while $1-P_{\mathrm{react}}$ is the fraction of trajectories which do
not lead to the final reaction event. Since all multipliers in result \eqref{react}
are less than unity, $P_{\mathrm{react}}$ exponentially decreases with the number
of stages in the reaction cascade.  
We also dwell some more on the asymptotic behaviour of expression
\eqref{react}, which harbours some subtle features. One observes that
$P_{\mathrm{react}}$ vanishes as soon as either of the $r_j$ becomes
equal to zero meaning that the terminal reaction event does not take
place at all. In this limit the corresponding target site becomes
point-like, such that it cannot be found by a search
process in a three-dimensional space -- this stage in a cascade cannot
be accomplished successfully.  When either of the $\rho_j$ becomes
infinitely large, $P_{\mathrm{react}}$ also vanishes, as it
should. For perfect reactions (all $\kappa_j \to \infty$) and $D_j
>0$, the probability of the eventual reaction depends only on the
geometrical parameters,
\begin{equation}
\label{react2}
P^{(\rm perfect)}_{\mathrm{react}}=\prod\limits_{j=1}^N\dfrac{r_j}{\rho_j},
\end{equation}
and is independent of the diffusion coefficients.  In this case,
one gets a very simple formula,  which is almost identical to \eqref{term1d}, 
\begin{equation}  \label{eq:Ht_3d_perfect}
H(t) = P^{(\rm perfect)}_{\mathrm{react}} \,\sqrt{\frac{T}{4\pi t^3}}\exp \left(-\frac{T}{4 t}\right).
\end{equation}
 For finite $\kappa_j$, however, $P_{\mathrm{react}}$ depends
explicitly on all diffusion coefficients. Naturally,
$P_{\mathrm{react}}$ also vanishes when either of the $\kappa_j$
equals zero, i.e., the corresponding intermediate stage within the
cascade is inhibited.  The above asymptotic behaviour is physically
plausible and intuitively clear.  Conversely, in the limit when either
of the diffusion coefficients $D_j$ vanishes, or, in contrast, tends
to infinity, expression
\eqref{react} produces a somewhat counter-intuitive behaviour: 
for $D_j \to 0$
the corresponding intermediate stage within the cascade cannot be
accomplished such that the actual $P_{\mathrm{react}}$ has to vanish,
while expression \eqref{react} does not.  In turn, when $D_j \to
\infty$, meaning that the transport within the corresponding stage
takes place instantaneously, the contribution of this stage has to be
equal to $1$, while expression \eqref{react} predicts that
$P_{\mathrm{react}} = 0$.  The point is that in the derivation of
equation \eqref{react} we relied on the diffusion equation with the
Robin boundary condition, which, in turn, is based on the assumption
that the characteristic times of diffusion and reaction attempts
are the same. Hence, the diffusion coefficient $D_j$ is linked to
$\kappa_j$ on the level of the microscopic parameters; in this setting,
$D_j$ is thus not an independent parameter and cannot be tuned
independently of $\kappa_j$ (while the latter, in contrast, can be
tuned independently of $D_j$ by varying the reaction
probability). This observation resolves the apparently controversial
behaviour of $P_{\rm react}$.  Extensions of the Robin boundary
condition were recently proposed in \cite{Grebenkov20}.

\begin{figure}
\centering
\includegraphics[width=8cm]{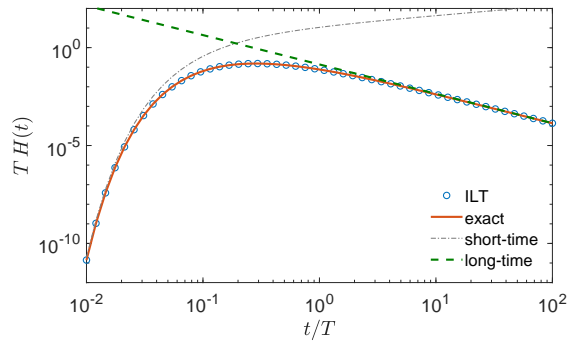}
\caption{ Rescaled
PDF $T H(t)$ as function of dimensionless time $t/T$ for a 2-stage cascade ($N=2$)  of imperfect reactions in three dimensions. Here, we choose
$r_1=1$, $\rho_1=2$, $D_1=1$ and $r_2=1.8$, $\rho_2=2$,
$D_2=0.4$,  while
$\kappa_1=10$ and $\kappa_2=1$. Circles represent the numerical Laplace
inversion (ILT) of equation \eqref{eq:Hp} with $N=2$ using the Talbot
algorithm; the solid line shows our exact 
result in \eqref{full_3d}; the dash-dotted line represents the short-time
asymptotic relation \eqref{eq:Ht_3D_short},
whereas the dashed  line indicates the long-time asymptotic relation
\eqref{eq:Ht_3D_long}.}
\label{fig2}
\end{figure}
In figure \ref{fig2} we depict our exact 
result in \eqref{full_3d} together with the asymptotic forms  \eqref{eq:Ht_3D_short} and \eqref{eq:Ht_3D_long}, and compare them against the numerical inversion of the Laplace transform (ILT) of the full expression \eqref{eq:Hp}. This comparison shows excellent agreement between  \eqref{full_3d} and the numerical inversion of the Laplace transform, 
and also indicates the limits in which the asymptotic forms  \eqref{eq:Ht_3D_short} and \eqref{eq:Ht_3D_long} describe accurately the tails of the PDF.

\section{Conclusion}
\label{conc}

We established a framework for molecular reaction cascades. Similar to
a relay race in human competitions, an incoming signal is here relayed
by stepwise reactions to a terminal target site. At each intermediate
target, activated through reaction with an incoming diffusive
messenger molecule, a new messenger molecule is produced, that in turn
needs to locate and react with the next reaction target, until the
final messenger molecule reacts with the last target. We obtained the
first-reaction time PDF $H(t)$ for the successful terminal reaction
event in a cascade of $N$ reaction steps in infinite one, two, and
three spatial dimensions along with the asymptotic behaviours in the
short and long time limits. We identified the time scale $T$ that is
characterised by the target-target distances and the diffusivities of
the respective messenger molecules. This specific form of $T$ defines
the most typical terminal reaction time and thus mirrors the
``geometry-control'' of direct trajectories from messenger release to
the associated reaction target, as observed earlier in one-step
reaction settings \cite{dist1,dist4} (see also \cite{direct} for a
``geometrical optics'' interpretation). Similar to the
L{\'e}vy-Smirnov form for the first-passage density we obtain an
exponential cutoff at short reaction times, modified by some
prefactors, and the long-time behaviour shows the expected power-law
scaling of the L{\'e}vy-Smirnov density. The dependence of the
first-reaction time PDF $H(t)$ on the system parameters depends on the
embedding dimension.

We established the reaction cascade framework here in an
infinite-space setting without boundaries. This scenario offers
relatively simple solutions, allowing us to spotlight the essential
dynamic features of reaction cascades.  
An intriguing question is what
will be different if the cascade (or parts thereof) will be placed in
finite reaction volumes. Infinite volumes may represent specific
cell-to-cell signals, e.g., from bacteria to predator cells such as
amoeba or to other bacteria in a growing biofilm.  From a broader perspective, 
 one may envisage their utility in
robotic search algorithms in which an encounter-propagation may also be of
relevance, as well as directed rumour spreading in societal models.
Finite volumes, in
turn, are relevant in scenarios such as internal signalling in cells.
Indeed, in finite volumes, additional time scales spanning
considerable time ranges will enter the specific form of $H(t)$ as
known for single-step reactions \cite{dist1,dist4}.  In finite
domains, the PDF of the first-reaction time with a single immobile
imperfect target generically consists of three temporal domains
delimited by characteristic time scales. These domains are: a
hump-like region for (relatively) short times, peaked at the most
probable reaction time and terminated at the instant of time when a
particle first engages with a boundary realising that it moves in a
bounded space. This short-time behaviour is followed by a plateau
regime, in which all the first-reaction times are equally
probable. Mathematically, such a regime emerges due to the gap between
the first and the second eigenvalues in the eigenvalue problem which
can be distinctly different and show a different dependence on the
parameters, in particular, on the chemical reactivity. Finally, the
plateau crosses over to an exponential decay, associated with the mean
first-reaction time.  Now, turning to the typical cellular domains
(with their geometrical peculiarities) and signal transduction
processes, which necessitate intermediate messengers operating between
the subdomains, we may expect that the PDF of the first-reaction time
at the terminal target site will have essentially the same functional
form. Clearly, in the long-$t$ limit one will find an exponential form
with the characteristic time scale which will be close to the mean terminal first-reaction time, equal to the 
sum of the mean
first-reaction times at the intermediate stages within the cascade of
reactions, 
 \eqref{1st}.
  Further on, at relatively short times, at which
the ``direct'' trajectories dominate, one will find either the
L{\'e}vy-Smirnov $t$-dependence for perfect reactions or the
$t$-dependence in our equations
\eqref{shorta} and \eqref{eq:Ht_short} for imperfect ones,  which we thus expect to be generic. Conversely, a careful analysis for particular, geometrically-relevant
settings is required to calculate the amplitudes in both asymptotic
laws and their dependence on the system parameters and also the
functional form of the eigenvalues, which define the duration of the
plateau region. Such an analysis, which is indispensable for
understanding the relative importance of different regimes, as indicated, e.g., 
by the fraction of reaction events happening at each stage, and
eventually, the functioning of the signal transduction processes, will
be presented elsewhere.

Apart from the above mentioned geometrical particularities of 
many realistic cascade processes, as exemplified, e.g., by a signal transduction in cellular environments,
our analytical description of such 
molecular relay races can be extended  further in several other directions.  In particular, upon an interaction of a given 
messenger
with a corresponding target site, often not a single but multiple 
messengers are released. This is an important aspect and its impact on 
the first-reaction times and their distribution for a search for a single target site
 has been rather 
extensively discussed within the recent 
years (see, e.g.,  \cite{N1,N2,N3,N4} and \cite{dist1a} and references therein).  Second, a target site
for a given messenger may not be unique, as we have supposed here,
but rather there will exist 
an array of such sites, each capable to launch a subsequent messenger. For instance, 
receptors on the cellular membrane may be present in sufficiently big amounts (see, e.g., \cite{alberts, snustad}).
Next, in biophysical applications, environments in which the particles move are typically extremely complex such that the latter 
interact with immobile obstacles of different kinds as well as  with
other mobile particle, which form a dense dynamical background but are not participating in reactions themselves. Such frenetic molecular crowding environments are known to 
entail a non-Gaussian dynamics, at least at short time scales (see, e.g., \cite{z1,Lanoiselee18,z2}), which will certainly affect 
the PDF of the terminal reaction time. Moreover, the diffusing molecules may (de-)polymerise on their way \cite{w1,w2,w3} or significantly shift their shape \cite{w4}, both effects leading to changes of their diffusivities as function of time.
Lastly, as interesting perspective consists in studying the effect
of the excluded volume 
of the target sites that was ignored in the present
work.  This question is apparently not very relevant for the signal transduction processes, for which
different messengers operate in different subdomains such that the target sites involved in previous and subsequent intermediate stages have no effect on the dynamics of a given messenger, but may become relevant in infinite systems.
In fact, as the particle searches for the $j$th target, the
other targets would typically present inert impermeable obstacles to
its motion.  When the targets are small and uniformly spread in the
medium, their hard-core exclusion effect is expected to be weak, at least at
long times.  In turn, as the particle starts the search at the $j$th
stage on the surface of the $(j-1)$th target, the hindering effect
might be notable at short times.  While exact solutions are not
available in such geometric configurations, asymptotic methods 
\cite{5,Lawley19,Ward06} and semi-analytical approaches
\cite{Chen09,Gordeliy09,Galanti16,Grebenkov19b,Grebenkov20b} can be applied.

\ack

The authors acknowledge helpful discussions with E. Sheval. DG acknowledges
the Alexander von Humboldt Foundation for support within a Bessel Prize award.
RM acknowledges support from the German Science Foundation (DFG grant ME
1535/12-1) as well as from the Foundation for Polish Science Fundacja na rzecz
Nauki Polskiej, FNP) through an Alexander von Humboldt Polish Honorary Research
Scholarship.

\appendix

\end{document}